\documentclass[proceedings]{JHEP3}
\usepackage{eurosym}
\usepackage{amssymb}
\usepackage{amsfonts}
\usepackage{amsmath,epsfig}
\usepackage{graphicx}
\usepackage{multirow}
\usepackage{float}
\usepackage{xcolor}
\newbox\mybox

\newcommand\fverb{\setbox\mybox=\hbox\bgroup\verb}
\newcommand\fverbdo{\egroup\medskip\noindent\fbox{\unhbox\mybox}\ }
\newcommand\fverbit{\egroup\item[\fbox{\unhbox\mybox}]}
\conference{'t Hooft-Polyakov monopoles in non-Hermitian QFT}
\abstract{We construct exact ’t Hooft-Polyakov monopole solutions in a non-Hermitian field theory with local non-Abelian SU(2) gauge symmetry
and a modified antilinear CPT symmetry. The solutions are obtained in a fourfold Bogomolny-Prasad-Sommerfield scaling limit giving rise to two different types of
monopole masses that saturate the lower energy bound. These two masses only coincide in the Hermitian limit and in the limit in which the symmetry breaking vacuum
tends to the trivial symmetry preserving vacuum. In the two theories corresponding to the two known Dyson maps these two masses are exchanged,
unlike the Higgs and the gauge masses, which remain the same in both theories. We identify three separate regions in parameter space bounded by different types
of exceptional points. In the first region the monopole masses are finite and tend both to zero at the boundary exceptional point, in the second the
monopole masses become complex and in the third only one of the monopole masses becomes zero at the boundary exceptional point, whereas the other tends to infinity.
We find a self-dual point in parameter space at which the gauge mass becomes exactly identical to the monopole mass.}

\title{'t Hooft-Polyakov monopoles in non-Hermitian quantum field theory}
\author{Andreas Fring and Takanobu Taira \\
Department of Mathematics, City, University of London,\\
Northampton Square, London EC1V 0HB, UK \\
E-mail: a.fring@city.ac.uk, takanobu.taira@city.ac.uk}

\input{tcilatex}

\begin{document}

\section{Introduction}
    Ever since Dirac demonstrated 89 years ago \cite{dirac31} that the postulate of the existence of magnetic monopoles
provides an explanation for the quantization of the electric charge, they have remained an appealing theoretical concept,
despite the fact that up to now magnetic monopoles have not been observed in nature. See \cite{mitsou2018quest,mavromatos2020} for a very recent
update on the experimental searches of magnetic monopoles in particle colliders and of cosmic origin that have been carried out, are currently performed
and on future plans. 

The concept of monopoles became an integral part of particle physics, notably of any theory aiming
at the formulation of grand unification of all fundamental forces, after ’t Hooft \cite{t1974magnetic} and Polyakov \cite{polyakov1974} noticed
that gauge theories almost inevitable contain monopole solutions.
Subsequently the mechanism for the emergence of monopoles, together with their properties and relations to the internal structures of particular
gauge theories have been extensively studied and many aspects are very well understood, see e.g. \cite{goddard1978,weinberg2012classical}. 
Here we are especially interested in the relation of their masses to the gauge particles and whether the intriguing features found in Hermitian theories also
hold in non-Hermitian versions. Especially if the property found by Montonen and Olive \cite{montonen1977magnetic} is still valid, that in non-Abelian gauge theories the soliton
solutions become equivalent to gauge massive fields in a dual theory.

Thus our aim is to study the properties of the monopole solutions in a non-Hermitian field theory with local non-Abelian gauge symmetry
and a modified antilinear $\mathcal{CPT}$ symmetry. We build on our previous investigations \cite{fring2019goldstone1,fring2019goldstone2,fring2020higgs}, and further elaborate on a particular model
studied in \cite{fring2020higgs} for which the Higgs masses have been identified in all PT-regimes. Variants of this model have also been investigated with different types of methods
in \cite{kings_eqm1,kings_eqm2,kings_abelian_higgs,kings_goldstone,kings_non-abelian_higgs,mannheim2019goldstone}.

We construct the monopole solutions in a fourfold Bogomolny-Prasad-Sommerfield (BPS) \cite{bogomol1976stability,prasad1975exact} scaling limit that has also been
successful in the Hermitian setting. We shall investigate the properties of these solutions and in particular establish whether the Montonen-Olive strong-weak
duality still holds.

Our manuscript is organised as follows: In section 2 we recall the non-Hermitian field theory previously studied in \cite{fring2020higgs} and set up the equations of motion whose
asymptotic solutions tend to the vacuum solutions. In sections 3 we discuss the Bogomolny energy bound for our theory. In section 4 we carry out the BPS-limit
in form of a fourfold scaling limit obtaining two different types of masses from the monopole solutions. In section 5 we discuss the properties of these masses
and identify physical regions in the parameter space. Our conclusions are stated in section 6.
\section{Soliton solution in a non-Hermitian model with $SU(2)$ gauge symmetry}
        Here we consider the non-Hermitian $SU(2)$ gauge theory with matter fields in the adjoint representation. This is a non-Hermitian extension of the Lagrangian studied in \cite{t1974magnetic,polyakov1996particle} which is known to possess monopole solutions with finite energies
        \begin{eqnarray}\label{adj_rep_action}
            S&=&\int d^4 x ~~ \frac{1}{2}Tr\left(D\phi_1\right)^2 +\frac{1}{2}Tr\left(D\phi_2\right)^2 -c_1 \frac{m_1^2}{2} Tr(\phi_1^2) + c_2 \frac{m_2^2}{2} Tr(\phi_2^2)\nonumber\\
            &&-i \mu^2 Tr(\phi_1 \phi_2)-\frac{g}{4}\left(Tr(\phi_1^2)\right)^2-\frac{1}{4}Tr(F^2).
        \end{eqnarray}
        Here we take $g,\mu\in\mathbb{R}$, $m_i \in\mathbb{R}$ and discrete values $c_i \in \{-1,1\}$. The two fields $\{\phi_i \}_{i=1,2}$ are Hermitian matrices $\phi_i (t,\Vec{x}) \equiv \phi_i^a (t,\Vec{x}) T^a$ where $\phi_i^a (t,\Vec{x})$ is a real-valued field. The three generators $\{T^a \}_{a=1,2,3}$ of $SU(2)$ in the adjoint representation are defined by three Hermitian matrices of the form $(T^a)_{bc} = -i \epsilon_{abc}$, satisfying the commutation relation $[T^a, T^b]=i\epsilon^{abc} T^c$. One can check that $Tr(T^a T^b)=2\delta ^{ab}$.
        The field strength tensor is defined as $F_{\mu\nu} = \partial_\mu A_\nu - \partial_\nu A_\mu -i e [A_\mu,A_\nu]$ where the gauge fields are $A_\mu  = A_\mu^a T^a$. 
        
        This action is invariant under the following local $SU(2)$ transformation of the matter fields and gauge fields
        \begin{eqnarray}
            \phi_i &\rightarrow &e^{i \alpha^a(x) T^a } \phi_i e^{-i \alpha^a(x) T^a },\\
            A_\mu &\rightarrow & e^{i \alpha^a(x) T^a }A_\mu  e^{-i \alpha^a(x) T^a }+\frac{1}{e}\partial_\mu \alpha^a(x) T^a .
        \end{eqnarray}
        This action is also symmetric under modified $\mathcal{CPT}$ symmetry where one of the matter fields $\phi_1$ transforms as scalar field and the matter field $\phi_2$ transforms as pseudo-scalar. This antilinear symmetry is an important indicator for the reality of the classical masses of the matter fields as extensively discussed in \cite{fring2019goldstone1,mannheim2019goldstone}
        \begin{equation}
           \mathcal{CPT}: \phi_1 (t,\Vec{x})\rightarrow  \phi_1 (-t,-\Vec{x}) ~,~~~\phi_2 (t,\Vec{x})\rightarrow  -\phi_2 (-t,-\Vec{x}) ~,~~~ i \rightarrow -i .
        \end{equation}
        The equations of motion for the fields $\phi_i$ and $A_\mu$ are
        \begin{equation}\label{equations of motion}
            \left(D_\mu D^\mu \phi_i\right)^a + \frac{1}{2}\frac{\delta V }{\delta \phi^a_i}=0 ~,~~~  D_\nu F^{\nu\mu}_a-e \epsilon_{abc}\phi^b_1 (D^\mu \phi)^c +e \epsilon_{abc}\phi^b_2 (D^\mu \phi)^c =0 .
        \end{equation}
        One of the problems often associated with non-Hermitian field theories is the incompatibility of the set of equations of motion \cite{kings_eqm1}. To overcome these issues we employed the Pseudo-Hermitian method which consists of mapping the non-Hermitian model to a Hermitian model through a similarity transformation. This is a common procedure used in the $\mathcal{PT}$ symmetric quantum mechanics \cite{ptbook,mostafazadeh2010pseudo}, and  in close analogy also their field-theoretic versions were studied for several different models \cite{bender2005dual,mannheim2019goldstone}. The similarity transformations for our model eq(\ref{adj_rep_action}) have already been introduced in \cite{fring2020higgs}
        \begin{equation}\label{similarity transformation}
            \eta_{\pm} = \prod_{a=1}^3 \exp\left(\pm\frac{\pi}{2}\int d^3 x \Pi^a_2 \phi^a_2 \right).
        \end{equation}
        The adjoint action of $\eta_\pm$ maps the complex action in eq(\ref{adj_rep_action}) into the following real action
        \begin{eqnarray}\label{real adj rep action}
            \eta_\pm S \eta^{-1}_\pm &=&\int d^4 x ~~ \frac{1}{2}Tr\left(D\phi_1\right)^2 -\frac{1}{2}Tr\left(D\phi_2\right)^2 -c_1 \frac{m_1^2}{2} Tr(\phi_1^2) - c_2 \frac{m_2^2}{2} Tr(\phi_2^2)\nonumber\\
            &&-c_3\mu^2 Tr(\phi_1 \phi_2)-\frac{g}{4}\left(Tr(\phi_1^2)\right)^2-\frac{1}{4}Tr(F^2)\nonumber\\
            &\equiv & \int d^4 x ~~ \frac{1}{2}Tr\left(D\phi_1\right)^2 -\frac{1}{2}Tr\left(D\phi_2\right)^2 - V -\frac{1}{4}Tr(F^2),
        \end{eqnarray}
        where the parameter $c_3$ takes the value $\pm 1$ for $\eta_\pm$ respectively. Notice that this model is very similar to those with the actions considered in \cite{t1974magnetic,polyakov1996particle}, but with second order coupling term $\mu^2 Tr(\phi_1 \phi_2)$ and negative sign in the kinetic term of $\phi_2$.
        
        Next one can use the simple scaling argument \cite{derrick_scaling_argument} to show that monopole solutions with finite energy require the monopole to asymptotically approach the vacuum solution $ V[\phi^0 ]=0$ (note that one can add a constant to the action so that this asymptotic condition is equivalent to $\delta V[\phi^0 ]=0$). The explicit values of the vacuum solutions $\phi_\alpha^0$ and $A_\mu^0$ are found by solving $\delta V =0$ and $D_\mu \phi_\alpha =0 $ \cite{general_solution_to_higgs_vacuum}
        \begin{eqnarray}\label{higgs vacuum}
            &(\phi_1^0)^a = \pm R \hat{r}^a \equiv h_1^{0\pm}\hat{r}^a~,~~~ (\phi_2^0)^a = \mp\frac{c_2 c_3\mu^2}{m_2^2} R \hat{r}^a\equiv h_2^{0\pm}\hat{r}^a  ,\nonumber\\
            &(A_i^0)^a =-\frac{1}{e}\epsilon^{abc}\hat{r}^b \partial_i \hat{r}^c + \hat{r}^a A_i = -\frac{1}{er}\epsilon^{iaj}\hat{r}^j+ \hat{r}^a A_i  ~,~~~(A_0^0)^a  =0,
        \end{eqnarray}
        where $r = (x,y,z)$, $R^2 \equiv (c_2 \mu^4 -c_1 m_1^2 m_2^2)/(2 g m_2^2)$ and overhat indicate the normalisation $\hat{r}=r/\sqrt{x^2 +y^2 +z^2}$. The $A_i$ are arbitrary functions of space-time.
        This vacuum solution is sometimes called the Higgs vacuum \cite{first_mensioned_higgs_vacuum} to distinguish it from the usual much simpler vacuum solution where $A^0_\mu =0$. The asymptotic condition can be written more explicitly if we consider the spherical ansatz
        \begin{equation}\label{Spherical ansatz}
            (\phi^{cl}_\alpha)^a (\Vec{x})=  h_\alpha (r) \hat{r}^a ~,~~~ (A_i^{cl})^a = \epsilon^{iaj}\hat{r}^j A(r)~,~~~ (A_0^{cl})^a =0,
        \end{equation}
        where the subscript $cl$ denotes the classical solutions to the equations of motion eq(\ref{equations of motion}).
        Here we are only considering the static ansatz to simplify our calculation, but one may of course also consider the time-dependent solution. For the monopole solution to have finite energy, we require the two matter fields of eq(\ref{Spherical ansatz}) to approach the vacuum solutions in eq(\ref{higgs vacuum}) at spacial infinity
        \begin{equation}\label{asymptotic condition}
            \lim_{r\rightarrow\infty} h_1 (r) =h_1^{0\pm}= \pm R~,~~~ \lim_{r\rightarrow\infty} h_2 (r) =h_2^{0\pm}= \mp \frac{c_2 c_3\mu^2}{m_2^2} R. 
        \end{equation}
        Also notice that at some fixed value of the radius $r$, the vacuum solutions $\phi_\alpha^0$ and monopole solutions $\phi_\alpha^{cl}$ both belongs to the 2-sphere in the field configuration space. For example, $\phi_1^0$ belong to the 2-sphere with radius $R$ because $(\phi_1^0)^2 = R^2$. Moreover, these vacuum and monopole solutions belong to the homotopy group $\pi^2 (S^2) = \mathbb{Z}$. It maps the 2-sphere in the space-time to 2-sphere in the field configuration space, which implies that there are $n\in \mathbb{Z}$ many topologically inequivalent solutions that can be explicitly written by replacing $\hat{r}^a$ in eq(\ref{higgs vacuum}), (\ref{Spherical ansatz}) with   
        \begin{equation}
            \hat{r}_n^a = \left(\begin{array}{c}
                \sin(\theta)\cos(n \varphi)  \\
                \sin(\theta)\sin(n \varphi) \\
                \cos(\theta)
            \end{array}\right).
        \end{equation}
        Since we require the monopole and vacuum solutions to smoothly deform into each other at spacial infinity, both solutions need to share the same integer $n$ usually referred to as the winding number. It is important to note that winding numbers of $\phi_1 $ and $\phi_2$ need to be equal to satisfy $D\phi_1 = D\phi_2 =0$ and therefore we will denote the winding numbers of $\phi_1$ and $\phi_2$ as $n$ collectively. If they are not equal we would have $D\phi_1 =0$ but $D\phi_2 \not=0$. Next, let us insert our ansatz eq(\ref{Spherical ansatz}) into the equations of motion eq(\ref{equations of motion}) by also making an explicit choice for $A(r)$
        \begin{equation}\label{t'Hooft Polyakov ansatz}
            (\phi_\alpha^{cl})^a = h_\alpha (r) \hat{r}_{n_{\alpha}}^a ~,~~~ (A_i^{cl})^a = \epsilon^{iaj}\hat{r}_n^j \left(\frac{u(r)-1}{e r}\right).
        \end{equation}
        This ansatz is more in line with the original ansatz given in \cite{bogomol1976stability,prasad1975exact}, compare to  eq(\ref{Spherical ansatz}).
        Inserting these expressions into the equations of motion eq(\ref{equations of motion}) we find 
        \begin{eqnarray}
            u^{''}(r)+\frac{u(r)\left[1-u^2 (r)\right]}{r^2} +\frac{ e^2 u(r)}{2} \left\{h_2^2 (r) - h_1^2 (r)\right\} &=&0,\label{t'Hooft Polyakov differential equations 1}\\
            h_1^{''}(r)+\frac{2h_1^{'} (r)}{r}-\frac{2 h_1(r)u^2(r)}{r^2} +g\left\{ c_1 \frac{m_1^2}{g} h_1 (r)+c_3 \frac{  \mu^2}{g} h_2 (r)+2 h_1^3  (r)\right\}&=&0,\label{t'Hooft Polyakov differential equations 2}\\
            h_2^{''}(r)+\frac{2h_2^{'} (r)}{r}-\frac{2 h_2(r)u^2(r)}{r^2} +c_2 m_2^2 \left\{ h_2 (r)+ c_3 \frac{\mu^2}{m_2^2} h_1 (r)\right\}&=&0 .\label{t'Hooft Polyakov differential equations 3}
        \end{eqnarray}
        Notice that these differential equations are similar to the ones discussed in \cite{bogomol1976stability,prasad1975exact}, but with the extra field $h_2$ and extra differential equation eq(\ref{t'Hooft Polyakov differential equations 3}). In the Hermitian model the exact solutions to the differential equations were found by taking the parameter limit called the BPS limit \cite{bogomol1976stability,prasad1975exact} where parameters in the theory are taken to zero while keeping the vacuum solution finite. Here we will follow the same procedure and take the parameter limit where quantities in the curly brackets of eq(\ref{t'Hooft Polyakov differential equations 2}) and (\ref{t'Hooft Polyakov differential equations 3}) vanish but keeping the vacuum solutions eq(\ref{higgs vacuum}) finite. We will see in section \ref{BPS limit section} that we also find the exact solutions in this limit. However, before we solve the differential equations, let us discuss the energy bound of the monopole.  
\section{The energy bound}\label{section:energy_bound}
        The energy of the monopole can be found by inserting the monopole solution into the corresponding Hamiltonian of eq(\ref{real adj rep action}).
        \begin{eqnarray}
            \mathfrak{h} &=& \int d^3 x ~~Tr\left(E^2\right) + Tr\left(B^2\right) +Tr\left\{(D_0 \phi_1)^2\right\}+Tr\left\{(D_i \phi_1)^2\right\}\nonumber\\
            &&-Tr\left\{(D_0 \phi_2)^2\right\}-Tr\left\{(D_i \phi_2)^2\right\}+V,
        \end{eqnarray}
        where $E,B$ are ${E^i}_a ={F_a}^{0i}$ , ${B^i}_a =- \frac{1}{2}\epsilon^{ijk}F_a^{ jk}$, $i,j,k \in \{1,2,3\}$. The gauge is fixed to be the radiation gauge (i.e ${A_a}^0 =0 , \partial_i {A_a}^i =0$). Notice that our monopole ansatz eq(\ref{t'Hooft Polyakov ansatz}) is static with no electric charge $E_i^a=0$ and therefore the above Hamiltonian simplifies to 
        \begin{eqnarray}
            E &=& \int d^3 x  ~~ Tr\left({B}^2\right) +Tr\left\{(D_i \phi_1)^2\right\}-Tr\left\{(D_i \phi_2)^2\right\}+V\nonumber\\
            &=& 2\int d^3 x  ~~ {B_i}^a {B_i}^a +(D_i \phi_1)^a (D_i \phi_1)^a -(D_i \phi_2)^a (D_i \phi_2)^a +\frac{1}{2}V.
        \end{eqnarray}
        Here, we simplified our expression by dropping the subscripts $A^{cl}_i \rightarrow A_i$ , $\phi_\alpha^{cl} \rightarrow \phi_\alpha$. We also keep in mind that these fields depend on the winding numbers $n\in \mathbb{Z}$. In the Hermitian model (i.e, when $\phi_2 =0$) one can rewrite the kinetic term as $B^2 + D\phi^2 = (B-D\phi)^2 + 2 BD\phi$ and find the lower bound to be $\int 2 BD\phi$. Here we will follow the similar procedure but we introduce some arbitrary constant $\alpha,\beta \in \mathbb{R}$ such that $B^2 = \alpha^2 B -\beta^2 B$ where $\alpha^2 -\beta^2 =1$. This will allow us to rewrite the above energy as
        \begin{eqnarray}\label{energy of the monopole}
            E &=& 2\int d^3 x ~~\alpha^2 \left\{{B_i}^a + \frac{1}{\alpha}(D_i \phi_1 )^a\right\}^2  -\beta^2 \left\{{B_i}^a + \frac{1}{\beta}(D_i \phi_2 )^a\right\}^2 \nonumber\\ 
            &&+ 2\left\{\alpha {B_i}^a (D_i \phi_1 )^a + \beta  {B_i}^a (D_i \phi_2 )^a \right\}+\frac{1}{2}V,
        \end{eqnarray}
        To proceed from here, we need to assume extra constraints on $\alpha$ and $\beta$ such that the following is true  
        \begin{eqnarray}
            \int d^3 x ~~\alpha^2 \left\{{B_i}^a + \frac{1}{\alpha}(D_i \phi_1 )^a\right\}^2  -\beta^2 \left\{{B_i}^a + \frac{1}{\beta}(D_i \phi_2 )^a\right\}^2&\geq &0, \label{boundary condition 1} \\
            \int d^3 x V &\geq& 0 . \label{boundary condition 2}
        \end{eqnarray}
        This allows us to write the lower bound on the energy as  
        \begin{eqnarray}
            E &\geq &2\int d^3 x  ~~ \left\{+\alpha {B_i}^a (D_i \phi_1 )^a + \beta {B_i}^a (D_i \phi_2 )^a \right\}\nonumber\\
            &=&2\int d^3 x~~  \alpha \left\{{B_i}^a \partial_i \phi_1^a + e {B_i}^a \epsilon^{abc}{A_i}^b \phi_1^c\right\}+ \beta \left\{{B_i}^a \partial_i \phi_2^a + e {B_i}^a \epsilon^{abc}{A_i}^b \phi_2^c\right\}\nonumber\\
            &=&2\int d^3 x~~  \alpha \left\{{B_i}^a \partial_i \phi_1^a + \left(-e\epsilon^{abc}{A_i}^b {B_i}^c\right)\phi_1^a\right\}+ \beta \left\{{B_i}^a \partial_i \phi_2^a +\left(-e\epsilon^{abc}{A_i}^b {B_i}^c\right)\phi_1^a \phi_2^c\right\}\nonumber\\
            &=&2\int d^3 x~~ \alpha \left\{{B_i}^a \partial_i \phi_1^a +\partial_i {B_i}^a \phi_1^a\right\}+\beta \left\{{B_i}^a \partial_i \phi_2^a + \partial_i {B_i}^a \phi_1^a\right\}\nonumber\\
            &=& 2\int d^3 x ~~ \alpha \partial_i \left({B_i}^a {\phi_1}^a\right)+\beta \partial_i \left({B_i}^a {\phi_2}^a\right)\nonumber\\
            &=&\lim_{r\rightarrow \infty}\left(2\alpha \int_{S_r} dS_i {B_i}^a {\phi_1}^a + 2\beta \int_{S_r} dS_i {B_i}^a {\phi_2}^a\right),
        \end{eqnarray}
        where in the fourth line we used $D_i B_i^a =0 $ which can be shown from the Bianchi identity $D_\mu \epsilon^{\mu\nu\rho\sigma}F^a_{\rho\sigma}=0$. The last line is obtained by using the Gauss theorem at some fixed value of the radius $r$. Since we are integrating over the 2-sphere with large radius, we can use the asymptotic conditions eq(\ref{asymptotic condition}) and replace the monopole solutions $\{\phi_\alpha^a, B_i^a\}$ with the Higgs vacuum $\{(\phi_\alpha^0)^a ,(B_i^0)^a\}$
        \begin{eqnarray}\label{energy bound}
            E&\geq & \left(2\alpha {\phi^0_1}^a  + 2\beta {\phi^0_2}^a\right)\lim_{r\rightarrow\infty}\int_{S_r} dS_i (B_i^0)^a \nonumber\\
            &=& \left(\pm2\alpha R \hat{r}^a_n \mp 2 \beta \frac{\mu^2}{m_2^2} R \hat{r}^a_n\right) \lim_{r\rightarrow\infty}\int_{S_r} dS_i (B_i^0)^a ,
        \end{eqnarray}
        where the upper and lower signs of the above energy correspond to the upper and lower signs of the vacuum solutions in eq(\ref{higgs vacuum}). The explicit value of $B_i^0$ can be obtained by inserting eq(\ref{higgs vacuum}) into 
        \begin{equation}
            (B_i^0)^a = - \frac{1}{2} {\epsilon_{i}}^{jk} \left(\partial_j A^0_k -\partial_k A^0_j + e A^0_j\times A^0_k \right)^a .
        \end{equation}
        After a lengthy calculation this expression can be simplified to $B_i^a = \hat{\phi_1^0}^{a} B_i = \hat{r}^a_{n} B_i$ where $B_i$ is defined as 
        \begin{equation}
            B_i \equiv -\frac{1}{2} \epsilon_{ijk} \left\{\partial^j A^k -\partial^k A^j +\frac{1}{e} \hat{r}_n\cdot \left(\partial^j \hat{r}_n \times \partial^k \hat{r}_n\right) \right\} .
        \end{equation}
        Notice that integrating the first term over the 2-sphere gives zero by Stokes theorem  $\int_S \partial \times A = \int_{\partial S} A =0$ where one can show that Stokes's theorem on closed surface gives zero by dividing the sphere into two open surfaces. The second term is a topological term which can be evaluated  \cite{Reference_for_winding_number} as
        \begin{equation}
            \int dS_i B_i = -\frac{4 \pi n}{e} .
        \end{equation}
        This is the magnetic charge of the monopole solutions. Notice that we could have chosen $B_i^a = \hat{\phi_2^0}^{a} B_i$ instead, which also leads to $\int dS_i B_i = -\frac{4 \pi n}{e}$ since we require winding numbers of $\phi_1$ and $\phi_2$ to be equal. 
        Finally we find our lower bound of the monopole energy 
        \begin{eqnarray}\label{energy bound of the monopole}
            E &\geq & 2 R \left(\pm\alpha \mp\beta \frac{c_2 c_3 \mu^2}{m_2^2}\right)\hat{r}^a_n \hat{r}^a_{n} \left(\frac{-4\pi n }{e}\right) = \frac{-8\pi n }{e}R \left(\pm\alpha\mp\beta \frac{c_2 c_3\mu^2}{m_2^2}\right) .
        \end{eqnarray}
        Notice that we have some freedom to choose $\alpha ,\beta\in\mathbb{R}$ as long as our initial assumptions eq(\ref{boundary condition 1}) are satisfied. We will see in the next section that we can take a parameter limit of our model which saturates the above inequality and gives an exact values to $\alpha$ and $\beta$. 
\section{The fourfold BPS scaling limit}\label{BPS limit section}
    Our main goal is now to solve the coupled differential equations eq(\ref{t'Hooft Polyakov differential equations 1})-(\ref{t'Hooft Polyakov differential equations 3}). Prasad, Sommerfield and Bogomolny \cite{prasad1975exact,bogomol1976stability} managed to find the exact solution by taking the parameter limit which simplifies the differential equations. The multiple scaling limit is taken so that all the parameters of the model tend to zero with some combinations of the parameter remaining finite. The combinations are taken so that the vacuum solutions stay finite in this limit. Inspired by this, we will take here a fourfold scaling limit
    \begin{equation}\label{modified BPS limit}
        g,m_1,m_2,\mu\rightarrow 0 ~,~~~ \frac{m_1^2}{g}<\infty ~,~~~ \frac{\mu^2}{g}<\infty ~,~~~ \frac{\mu^2}{m_2^2}<\infty .
    \end{equation}
    This will ensure that the vacuum solutions eq(\ref{higgs vacuum}) stays finite, but crucially the curly bracket parts in eq(\ref{t'Hooft Polyakov differential equations 2}), (\ref{t'Hooft Polyakov differential equations 3}) vanish. There is a physical motivation for this limit in which the mass ratio of the Higgs and gauge mass are taken to be zero (i.e $m_{\text{Higgs}}<< m_g$) as described in \cite{physical_motivation_for_BPS}. We will see in the next section that the same type of behaviour is present in our model, hence justifying eq(\ref{modified BPS limit}).  The resulting set of differential equations, after taking the BPS limit is similar to the ones considered in \cite{prasad1975exact,bogomol1976stability} with the slightly different quadratic term in eq(\ref{t'Hooft Polyakov differential equations 1}). It is natural to consider a similar ansatz as given in \cite{prasad1975exact,bogomol1976stability} 
    \begin{eqnarray}
        u(r) &=& \frac{evr}{\sinh{(evr)}},\label{BPS ansatz 1}\\
        h_1 (r)&=&  -\alpha \left(v \coth{(evr)}-\frac{1}{er}\right)\equiv  -\alpha f(r),\label{BPS ansatz 2}\\
        h_2 (r)&=&-\beta \left(v \coth{(evr)}-\frac{1}{er}\right)\equiv -\beta f(r).\label{BPS ansatz 3}
    \end{eqnarray}
   where $\alpha,\beta \in \mathbb{R}$ were introduced in section (\ref{section:energy_bound}) and $f(r)\equiv \left\{v\coth\left(evr\right)-\frac{1}{er}\right\}$. One can check that this ansatz indeed satisfies differential equations eq(\ref{t'Hooft Polyakov differential equations 1})-(\ref{t'Hooft Polyakov differential equations 3}) in the BPS limit. We have decided to put a prefactor $\alpha$ and $\beta$ in front of eq(\ref{BPS ansatz 2}),(\ref{BPS ansatz 3}) to satisfy the differential equation eq(\ref{t'Hooft Polyakov differential equations 1}). Note that if we take $\alpha=1$ we get exactly the same as given in \cite{bogomol1976stability,prasad1975exact}, which is known to satisfy the first order differential equation called Bogomolny equation $B_i -D_i \phi =0$. The ansatz eq(\ref{BPS ansatz 1})-(\ref{BPS ansatz 3}) only differs from the ones given in \cite{bogomol1976stability,prasad1975exact} by the prefactors $\alpha$ and $\beta$, and therefore our ansatz should satisfy Bogomolny equation with the appropriate prefactor to cancel the prefactor in eq(\ref{BPS ansatz 2}),(\ref{BPS ansatz 3})    
    \begin{eqnarray}
        B_i^b + \frac{1}{ \alpha} (D_i \phi_1)^b &=&0,\\
        B_i^b + \frac{1}{\beta} (D_i \phi_2)^b&=&0,
    \end{eqnarray}
    where $\phi_\alpha \equiv h_\alpha (r) \hat{r}_{n}$. 
    If we compare these equations to the terms appearing in the energy of the monopole eq(\ref{energy of the monopole}), then we can saturate the inequality in eq(\ref{energy bound of the monopole}) by
    \begin{eqnarray}\label{saturated_energy}
        E[\phi_1 ,\phi_2]=\frac{-8\pi n R}{e} \left(\pm\alpha\mp\beta \frac{c_2 c_3\mu^2}{m_2^2}\right),
    \end{eqnarray}
    where upper and lower signs correspond to the vacuum solutions eq(\ref{higgs vacuum}). Note that this equality is true up to a constant given by $\int d^3 x V$ which can be removed by introducing an appropriate constant in the action eq(\ref{real adj rep action}). We can calculate the explicit forms of $\alpha$ and $\beta$ by comparing the asymptotic conditions in eq(\ref{asymptotic condition})
    \begin{equation}\label{matching_asymptotic_condition_1}
        \lim_{r\rightarrow \infty}h_1^\pm =h_1^{0\pm}= \pm R ~,~~~ \lim_{r\rightarrow \infty}h_2^\pm =h_2^{0\pm}= \mp \frac{c_2 c_3\mu^2}{m_2^2}R,
    \end{equation}
    with the asymptotic values of eq(\ref{BPS ansatz 1})-(\ref{BPS ansatz 3})
    \begin{equation}\label{matching_asymptotic_condition_2}
        \lim_{r\rightarrow \infty} u(r) =0 ~,~~~ \lim_{r\rightarrow \infty} h_1^\pm(r) =-\alpha v ~,~~~\lim_{r\rightarrow \infty} h_2^\pm(r) =-\beta v.
    \end{equation}
   Comparing the eq(\ref{matching_asymptotic_condition_1}) and eq(\ref{matching_asymptotic_condition_2}) we find the algebraic equations for $\alpha$ and $\beta$. Using $\alpha^2 - \beta^2 =1$ and assuming $m_2^4 \geq \mu^4$, we find the two set of real solutions 
    \begin{equation}\label{explict values of a and v}
        \alpha = (\pm)\frac{m_2^2}{l}~,~~~ v=\mp(\pm) \frac{R l}{m_2^2}~,~~~\beta =\mp (\pm)\frac{c_2 c_3 \mu^2}{l} ,
    \end{equation}
    where $l=\sqrt{m_2^4 -\mu^4}$. The plus-minus signs in the brackets correspond to the two possible solutions to the algebraic equation of $\alpha$. These need to be distinguished from the upper and lower signs of $v$ and $\beta$ which correspond to the vacuums solutions eq(\ref{higgs vacuum}). Inserting the explicit values of $\alpha$ and $\beta$ to the energy eq(\ref{saturated_energy}) we find
    \begin{equation}\label{all_possible_energies}
        E [\phi_1 ,\phi_2]\equiv (\pm)\frac{8\pi n R}{ e m_2^2} \left(\frac{\mp m_2^4 - \mu^4}{l}\right),
    \end{equation}
    where we observe two distinct energies for two vacuum solutions eq(\ref{higgs vacuum}). Notice that we need to take extra care when choosing the plus or minus sign in the brackets as wrong choices lead to negative energies.  To see this explicitly let us consider the monopole solutions with vacuum solutions $\{h_1^{0+} , h_2^{0+}\}$ as boundary conditions. The corresponding energy is 
    \begin{equation}
        E = (\pm) \frac{-8\pi n R}{ e m_2^2} \left(\frac{ m_2^4 + \mu^4}{l}\right).
    \end{equation}
    Then notice that the upper sign in the brackets leads to the positive energy only when $n<0$ and the lower sign leads to the positive energy when $n>0$. This allows us to choose the right sign for the $\alpha$ in eq(\ref{explict values of a and v}). This implies that the monopole solutions eq(\ref{BPS ansatz 1})-(\ref{BPS ansatz 3}) takes the following forms
    \begin{equation}
        \begin{array}{c|c|c|c}
            \text{Winding} &\{h_1 ,h_2 \}&u&\text{Mass} \\\hline
            n>0 &\left\{\frac{m_2^2}{l} f(r) , -\frac{c_2 c_3 \mu^2}{l}f(r)\right\} &\frac{eRlr}{m_2^2\sinh{(eRlr/m_2^2)}}&\frac{8\pi n R}{e m_2^2}\left(\frac{ m_2^4 +  \mu^4}{l}\right) \\\hline
            n<0 &\left\{-\frac{m_2^2}{l} f(r) , \frac{c_2 c_3 \mu^2}{l}f(r)\right\} &\frac{-eRlr}{m_2^2\sinh{(-eRlr/m_2^2)}}&\frac{8\pi n R}{e m_2^2}\left(\frac{ m_2^4 + \mu^4}{l}\right) 
        \end{array}
    \end{equation}
    where at rest the mass is equivalent to the energy of the corresponding monopole solutions. We can repeat the same analysis for the case when the monopole solutions asymptotically approach $\{h_1^{0-} , h_2^{0-}\}$ and find the different set of monopole solutions. We summarise here all possible solutions and its corresponding masses
    \begin{equation}
        \begin{array}{c|c|c|c|c}
            \text{Winding} &\{h_1^0 , h_2^0\}&\{h_1 ,h_2 \}&u&\text{Mass} \\\hline
            n>0 &\{h_1^{0\pm} , h_2^{0\pm}\}&\left\{\frac{m_2^2}{l} f(r) , \mp \frac{c_2 c_3 \mu^2}{l}f(r)\right\} &\frac{\pm eRlr}{m_2^2\sinh{(\pm eRlr/m_2^2)}}&M_\pm\\\hline
            n<0 &\{h_1^{0\pm} , h_2^{0\pm}\}&\left\{-\frac{m_2^2}{l} f(r) , \pm\frac{c_2 c_3 \mu^2}{l}f(r)\right\} &\frac{\mp eRlr}{m_2^2\sinh{(\mp eRlr/m_2^2)}}&M_\pm
        \end{array}
    \end{equation}
    Where $M_{\pm} = 8\pi |n| R (m_2^4 \pm \mu^4) / e m_2^2 l\geq 0$. We can combine these solutions and find
    \begin{equation}\label{Monopole solutions}
        \begin{array}{c|c|c|c}
           \{h_1^0 ,h_2^0\}  &\{h_1,h_2\} &u& \text{Mass} \\\hline
             \{h_1^{0\pm} ,h_2^{0\pm}\}&\left\{\text{Sign}(n)\frac{m_2^2}{l}f(r) ~,~~ \text{Sign}(n)\frac{\mp c_2 c_3\mu^2}{l}f(r)\} \right\} &\frac{\pm\text{Sign}(n) eRlr}{m_2^2\sinh{(\pm\text{Sign}(n) eRlr/m_2^2)}}& M_{\pm}
        \end{array}
    \end{equation}
    So we find two monopoles with two distinct masses characterised by the upper and lower signs of the vacuum solutions eq(\ref{higgs vacuum}) which can be interchanged by choosing opposite sign for $c_2 c_3$. This is because changing the sign of $c_2 c_3$ exchanges the vacuum solutions eq(\ref{higgs vacuum}), resulting in swapping of the monopole solutions. Note that the two masses and solutions only coincide in the Hermitian limit $\mu=0$. 
\section{Monopole mass and physical region}
    In this section, we will compare the monopoles masses with Higgs and massive gauge masses. We will analyse different versions of the model characterised by choosing different values of $c_i$. It was found in \cite{fring2020higgs,kings_non-abelian_higgs} that mapping the theories with respect to two different Dyson maps eq(\ref{similarity transformation}) does not effect the Higgs and gauge masses as they depended on $\mu^4$. Here we find that different Dyson maps corresponds to exchanging the two monopole solutions eq(\ref{Monopole solutions}). 
    \subsection{Higgs mass and exceptional points}\label{Section: Higgs recap}
        Let us begin by recalling the results from our previous work \cite{fring2020higgs}. The Higgs masses squared and gauge mass of our model defined in eq(\ref{adj_rep_action}) are 
        \begin{equation}\label{higgs mass}
            m_0^2 = -c_2 \frac{m_2^4 - \mu^4}{m_2^2} ~,~~~ m_{\pm}^2 = K\pm \sqrt{K^2 - 2 c_1 c_2 m_1^2 m_2^2 + 2 \mu^4}~,~~~ m_g = e\frac{|R l|}{m_2^2},
        \end{equation}
        where $K = -c_1 m_1^2 -c_2 \frac{m_2^2}{2} +\frac{3 \mu^4}{2 c_2 m_2^2}$. We recall that for some values of parameters, our masses can become complex, which is a common feature of non-Hermitian theories. The region where all Higgs masses are real were investigated in \cite{fring2020higgs,kings_non-abelian_higgs}. There are two non-overlapping physical regions for $c_1 = c_2 =\pm 1$ in our model eq(\ref{adj_rep_action}), which are equivalent to $-c_1 = c_2 =\pm 1$ in \cite{fring2020higgs}. Notice that in the BPS limit we have $m_0 = m_\pm =0$, but $m_g$ and $M_\pm$ stays finite, such that the ratios $m_{\text{Higgs}}/m_g$ vanish in the BPS limit. This is in line with the Hermitian case \cite{physical_motivation_for_BPS}, providing the physical interpretation $m_{Higgs}<<m_g$ for the BPS limit.
    
        We have observed that some boundaries of the physical region admit interesting behaviour where the gauge masses vanish. In particular, there are two distinct points (boundaries) called zero exceptional points of type $I$ and $II$ where the gauge masses vanish in each case. At the type $II$ zero exceptional points, the vanishing of gauge masses is not surprising because the vacuum manifold collapse to a point and the broken symmetry is restored. However, at the type $I$ zero exceptional point the vacuum manifold is finite and therefore the symmetry is still broken, nonetheless the gauge mass still loses its mass. The mass matrix for Higgs fields is also non-diagonalisable at this point, indicating that this point is a novel feature of the non-Hermitian theory. The type $I$ zero exceptional point which we will call 0EP, occurs when $m_2^4 = \mu^4$. 
    \subsection{Mass swapping of monopoles with similarity transformations}
        The monopole solutions of the type eq(\ref{BPS ansatz 1})-(\ref{BPS ansatz 3}) can not exist in the physical region $c_1 =c_2 =1$. To see this, recall that we assumed $m_2^4 - \mu^4 \geq 0$ in order to find the solutions to $\alpha$ and $\beta$ in eq(\ref{explict values of a and v}).
        Since we want our Higgs masses to be positive, we require $m_2^4 - \mu^4 \leq 0$ for $c_1 =c_2 =1$ and $m_2^4 - \mu^4 \geq 0$ for $c_1 =c_2 =-1$.
        Therefore we will only find the real solutions to $\alpha$ and $\beta$ in eq(\ref{explict values of a and v}) when $c_1 = c_2 =-1$. This implies that the monopole solutions of type eq(\ref{BPS ansatz 1})-(\ref{BPS ansatz 3}) can only exist for the case $c_1 =c_2 =-1$. This restricts our monopole solutions eq(\ref{Monopole solutions}) to be 
        \begin{equation}\label{final_monopole_solution}
        \begin{array}{c|c}
            \{h_1,h_2\} & \text{Mass} \\\hline
            \left\{\frac{m_2^2}{l} f(r) ,c_3\frac{\mu^2}{l} f(r) \right\} & M_+=\frac{8\pi R}{e m_2^2}\left(\frac{m_2^4 + \mu^4}{l}\right)\\\hline
            \left\{\frac{m_2^2}{l} f(r) ,-c_3\frac{\mu^2}{l} f(r) \right\} & M_- = \frac{8\pi R}{e m_2^2}l
        \end{array}
        \end{equation}
       here we took the winding number $n=1$. Let us denote the solutions with the same signs to be $(+,+)$ solution and the opposite signs to be $(+,-)$ solution. Then for example, if $c_3 =1$ then the first solution of above table is a $(+,-)$ solution. In summary we have
        \begin{equation}\label{final monopole masses}
            \begin{array}{c|c|c}
                \text{Solution} &\text{Mass when $c_3=1$} &\text{Mass when $c_3=-1$} \\\hline
                (+,+) \text{ solution} & M_+ &M_-\\\hline
                (+,-) \text{ solution}&M_- & M_+
            \end{array}
        \end{equation}
        The Higgs masses and gauge masses are independent of the sign of the non-Hermitian coupling $c_3$, but here we see that the monopole masses depend on the signs of the non-Hermitian coupling. This implies that the monopole mass depends on the similarity transformation in eq(\ref{similarity transformation}), where choosing $\eta_{\pm}$ corresponds to $c_3 = \pm 1$.  So we observe that the different monopole masses, resulting from different solutions, can be exchanged by using different similarity transformations. 
        
        We started with the four different possible theories $S_{c_1 c_2}$ in eq(\ref{adj_rep_action}) which can be transformed into many different theories under appropriate similarity transformations. Here we considered two similarity transformations $\eta_\pm$ which gives 8 possible theories $\eta_\pm S_{c_1 c_2}\eta^{-1}_\pm \equiv \mathfrak{s}_{c_1 c_2\pm}$. We have observed in \cite{fring2020higgs} that only $\mathfrak{s}_{++\pm}$ and $\mathfrak{s}_{--\pm}$ admit positive gauge masses. This is somewhat reminiscent of how $\phi^4$ theory can only admit symmetry breaking when the sign in the mass term is opposite to the sign in the $\phi^4$ term. If we focus on the $(+,+)$ solution then it appears at first sight that the two theories $\mathfrak{s}_{--+} $ and $\mathfrak{s}_{-- -}$ are inequivalent as the mass of the $(+,+)$ solution changes. However, notice that the masses are swapped between $(+,+)$ and $(+,-)$ solutions, therefore the two theories $\mathfrak{s}_{-- +} $ and $\mathfrak{s}_{-- -}$ are in fact equivalent with respect to the internal symmetry of exchanging the two monopole solutions. This internal symmetry can be seen in eq(\ref{final_monopole_solution}) where choosing the different Dyson map (i.e different values for $c_3$) leads to two monopole solutions to swap. Let us summaries this 
        \begin{equation}
            \begin{array}{c|c|c|c} 
                 \text{Theories}&\text{Gauge mass} &\text{Mass of }(++)\text{ solution}&\text{Mass of }(+-)\text{ solution} \\\hline
                 \mathfrak{s}_{++\pm}&m_g &\text{do not exist}&\text{do not exist}\\\hline
                 \mathfrak{s}_{--+}&m_g&M_+&M_-\\\hline
                 \mathfrak{s}_{---}&m_g&M_-&M_+
            \end{array}
        \end{equation}
        \subsection{Finite energy condition violation}
            As discussed in section \ref{Section: Higgs recap}, the gauge mass vanishes at the zero exceptional point $\mu^4 /m_2^4=1$. We notice that in this parameter limit our two monopole masses eq(\ref{final monopole masses}) vanish or diverge 
            \begin{equation}
                 \lim_{\frac{\mu^4}{m_2^4}\rightarrow 1} \frac{8\pi |n|R}{e m_2^2}l =0 ~,~~~\lim_{\frac{\mu^4}{m_2^4}\rightarrow 1} \frac{8\pi |n|R}{e m_2^2}\left(\frac{m_2^4 + \mu^4}{l}\right) \rightarrow \infty.
            \end{equation}
            We have taken here $R=\sqrt{(m_1^2 m_2^2-\mu^4)/2gm_2^2}$ since our monopole solution can only exist when $c_1=c_2=-1$. From now on we keep this choice and take $c_1=c_2=-1$.
            The monopole solutions eq(\ref{final_monopole_solution}) also diverge at the zero exceptional point. This is a similar effect to how Higgs fields are no longer identifiable at the zero exceptional point because they diverge and the Hamiltonian is no longer diagonalisable. Let us denote the two monopole masses as $M_- \equiv M_{cov}$ and $M_+ \equiv M_{div}$ for converging and diverging masses at 0EP limit. The converging mass can be written in terms of the gauge mass as 
            \begin{equation}
                M_{cov}=\frac{8 \pi |n|}{e} \frac{R l}{m_2^2} = \frac{8 \pi |n|}{e^2} m_g.
            \end{equation}
            Therefore we see that only one of the monopole masses satisfies the  Montonen-Olive duality \cite{montonen1977magnetic}, whereas the other mass does not satisfy this duality due to an additional factor $(m_2^4 +\mu^4)/(m_2^4 -\mu^4)$. 
            
            Let us rewrite the monopole and gauge masses in terms of the finite quantities in the BPS limit eq(\ref{modified BPS limit}) \begin{eqnarray}
                m_g &=& \frac{e}{\sqrt{2}} \sqrt{(X-YZ)(1-Z^2)},\label{gauge_mass}\\
                M_{cov}&=& \frac{8\pi |n|}{e} \sqrt{(X-YZ)(1-Z^2)},\label{cov_mono_mass}\\
                M_{div}&=&\frac{8\pi |n|}{e}\sqrt{X-YZ} \frac{1+Z^2}{\sqrt{1-Z^2}},\label{div_mono_mass}
            \end{eqnarray}
            where $m_1^2/g \equiv X, \mu^2/g\equiv Y ,\mu^2/m_2^2\equiv Z$. Therefore the 0EP corresponds to $Z=1$. One can view the parameters $Y$ an $Z$ as a measure of how strong the non-Hermitian term is coupled to the Hermitian theory. Notice that two monopole masses eq(\ref{cov_mono_mass}),(\ref{div_mono_mass}) are real and positive when the gauge mass eq(\ref{gauge_mass}) is real and positive. This implies that the physical region of monopole masses and gauge mass coincide when $c_1 = c_2 =-1$ and $Z\not=1$. Moreover, monopole masses and Higgs masses eq(\ref{higgs mass}) are both real and positive when $c_1 = c_2 =-1$ and $Z\not=1$. This is because the physical region of Higgs masses is a subset of the physical region of the gauge mass as seen in \cite{fring2020higgs}.
            
            Let us plot eq(\ref{gauge_mass})-(\ref{div_mono_mass}) for $n=1$, $X>Y$ (i.e $m_1^2 > \mu^2$) with a weak and a strong coupling $e=2, e=10$, respectively. 
            \begin{figure}[H]
                \centering
                \includegraphics[scale=0.5]{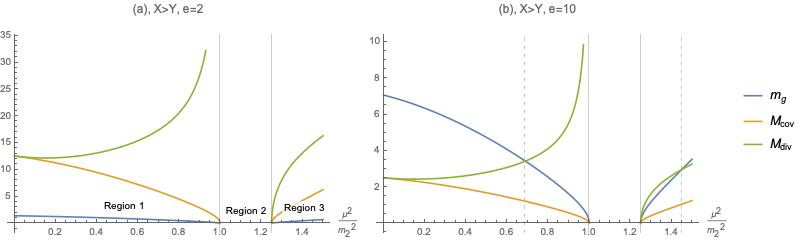}
                \caption{Both panels are plotted for $X=1, Y=0.8, |n|=1$ with different values for $e$. The dashed lines in panel (b) indicates the self-dual points where the gauge mass and the monopole mass coincide at $Z_0^- = 0.689 , Z_0^+ = 1.45$.}
                \label{fig:X bigger than Y}
            \end{figure} 
            The three regions are separated by the two types of exceptional points discussed in section \ref{Section: Higgs recap}. Region 1 is bounded between the Hermitian limit $Z\rightarrow 0$ and the above 0EP of type $I$ at $Z=1$, where the vacuum manifold stays finite. We see here that one of the monopole masses, $M_{div}$, diverges at the 0EP. This violates our initial assumption eq(\ref{asymptotic condition}) of the finite energy of the monopole. The reason for this is because the monopole solution eq(\ref{final_monopole_solution}) is ill-defined at the 0EP, therefore, one can not continuously deform the monopole solutions to the vacuum solutions eq(\ref{higgs vacuum}). In fact, the finite energy condition eq(\ref{asymptotic condition}) is also violated in the region 2 as the monopole solutions are complex, but the vacuum solutions are real and therefore they can not continuously be deformed into each other by taking $r\rightarrow \infty$. In region 3, the finite energy condition is restored as both the monopole and the vacuum solutions are complex. In summary:
            \begin{equation}
                \begin{array}{c|c|c|c}
                    &\text{Region 1}&\text{Region 2}&\text{Region 3}\\\hline
                    \text{Monopole} &\text{Real} &\text{Complex}&\text{Complex} \\\hline
                    \text{Vacuum} &\text{Real} &\text{Real}&\text{Complex}
                \end{array}
            \end{equation}
            In the strong coupling case $e=10$, we find self-dual points, $Z_0^\pm$, at which the gauge mass $m_g$ and the monopole mass $M_{div}$ become equal and interchange their relative size in both regions 1 and 3. This phenomena only occurs when the gauge mass is bigger than the monopole mass in the Hermitian limit $Z\rightarrow 0$. To see this, consider the monopole masses when $Z=0$ $M_0 = M_{div}|_{Z=0}=M_{cov}|_{Z=0}$. Then one can show that $m_g < M_0$ implies $e^2 < 8 \sqrt{2}|n|\pi$.  On the other hand, solving $m_g = M_{div}$ one finds 
            \begin{equation}
                (Z_0^{\pm })^2 = \frac{e^2 \pm 8\sqrt{2}|n|\pi}{e^2 \mp 8\sqrt{2}|n|\pi}> 0 \implies e^2 > 8 \sqrt{2}|n|\pi .
            \end{equation}
            Therefore $m_g < M_0$ contradict $(Z_0^{\pm })^2 >0$. Note that, the existence of  the self-dual points $Z_0^\pm$ only depends on the values of $e$ and $|n|$ and not on $X$ and $Y$. 
            
            Next let us consider the $X<Y$(i.e $m_1^2 < \mu^2$) case 
            \begin{figure}[H]
                \centering
                \includegraphics[scale=0.5]{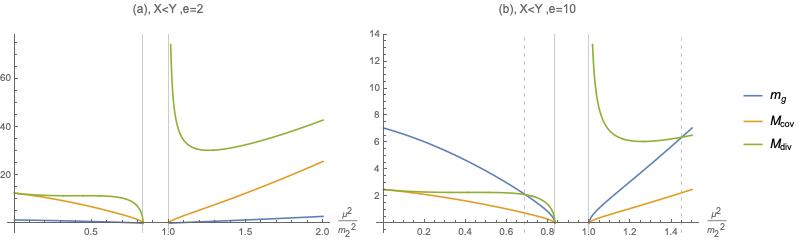}
                \caption{Both panels are plotted for $X=1, Y=1.2$ with different values for $e$. The dashed lines in panel (b) indicates the self-dual points where the gauge mass and the monopole mass coincide at $Z_0^- = 0.689 , Z_0^+ = 1.45$.}
                \label{fig:X smaller than Y}
            \end{figure}
            This case is equivalent to the $X>Y$ but with regions 1 and 3  exchanged. This means that the behaviour observed in the strongly non-Hermitian region in $X>Y$ of figure \ref{fig:X bigger than Y} is equivalent to the behaviour observed in the weakly non-Hermitian region in $X<Y$ of figure \ref{fig:X smaller than Y}.  
            
            Finally let us look at the $X=Y$ (i.e $m_1^2 = \mu^2$) case 
            \begin{figure}[H]
                \centering
                \includegraphics[scale=0.5]{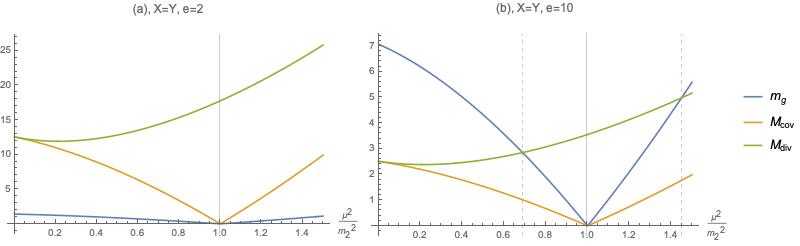}
                \caption{Both panels are plotted for $X=1, Y=1$ with different values of $e$. The dashed lines in panel (b) indicates the self-dual points where the gauge mass and the monopole mass coincide at $Z_0^- = 0.689 , Z_0^+ = 1.45$.}
                \label{fig:X is equal to Y}
            \end{figure}
            We see that the regions 2 collapses as the two boundaries coincide.  
            The gauge and monopole masses $m_g$ and $M_{cov}$ behave similarly to the $X>Y$ case but we see an interesting behaviour of the monopole mass $M_{div}$ at the 0EP. As we discussed in previous cases, the finite energy condition fails at the 0EP therefore one would expect to find unbounded energy. However, we observe that the monopole mass $M_{div}$ is finite even at the 0EP. This can be verified by looking at the asymptotic value of $M_{div}$ at $Z=1$
            \begin{equation}
                \lim_{Z\rightarrow 1}\left(\left.M_{div}\right|_{X=Y}\right)=\frac{8 \pi \sqrt{X}}{e}\lim_{Z\rightarrow 1}\left(\frac{\sqrt{1-Z}(1+Z^2)}{\sqrt{1-Z^2}}\right) = \frac{8\pi \sqrt{X}}{e}\sqrt{2} .
            \end{equation}
            At the 0EP the monopole solutions (\ref{final_monopole_solution}) are ill-defined and violate the finite energy condition eq(\ref{asymptotic condition}). However for $X=Y=1$ they remain finite.
            
            In the three cases we considered above, we see that the monopole mass can reach infinity with finite values of $Z$. For example, the region $1$ of the figure $1$ can not be physical as $M_{div}$ diverges at $Z=1$. We see that the only physical region (i.e no diverging mass with finite values of $Z$) is the region 1 of the case $X<Y$ and $X=Y$.

\section{Conclusion}
    We found and analysed the t'Hooft-Polyakov monopole in the non-Hermitian model possessing $SU(2)$ symmetry and anti-linear $\mathcal{CPT}$ symmetry using the pseudo-Hermitian approach. Following the procedure outlined in \cite{bogomol1976stability,prasad1975exact} we have found two exact monopole solutions eq(\ref{final_monopole_solution}) in the BPS limit which saturated the lower bound of the energy. These monopole solutions can only exist in one of the two physical regions characterised in \cite{fring2020higgs} by $c_1$ and $c_2$. We have considered four theories $S_{c_1 ,c_2}$ eq(\ref{adj_rep_action}) which were mapped to 8 theories via two similarity transformations eq(\ref{similarity transformation}). These transformations were previously found to not affect the Higgs masses and gauge masses as they have no dependencies on the similarity transformation parameter $c_3$. Here we observed two distinct masses for each monopole solutions to be interchangeable under different similarity transformations, as they depend on $c_3$. 
    
    The behaviour of the gauge and monopole masses were investigated as function of the non-Hermitian coupling $Z=\mu^2 /m_2^2$ . Three disconnected regions were found in $Z\in(0,\infty)$ where finite energy condition failed in one of the region, resulting in complex energies. At one of the boundary (0EP of type $I$) of the region, the monopole mass diverged to infinity, signalling that the theory is unbounded at this boundary. However, the monopole mass can have finite non-zero value even at 0EP of type $I$ when two boundaries (0EP of type $I$ and $II$) coincide.    
    
    The issue of finite energy at the boundary requires further investigation as it is peculiar to observe a finite energy of the solution which is ill-defined at the boundary. It is also natural to wonder if we find different monopole solutions with distinct masses if we consider different similarity transformation. 
\bibliographystyle{phreport}
\bibliography{bib.bib}

\begin{thebibliography}{10}

\bibitem{dirac31}
P.~A.~M. Dirac,
\newblock Quantised singularities in the electromagnetic field,
\newblock Proc. of the Royal Soc. of London A {\bf 133}(821), 60--72 (1931).

\bibitem{mitsou2018quest}
V.~A. Mitsou,
\newblock The quest for magnetic monopoles--past, present and future,
\newblock Proceedings of Science , 188 (2018).

\bibitem{mavromatos2020}
N.~E. Mavromatos and V.~A. Mitsou,
\newblock Magnetic monopoles revisited: Models and searches at colliders and in
  the Cosmos,
\newblock arXiv preprint arXiv:2005.05100  (2020).

\bibitem{t1974magnetic}
G.~'t~Hooft,
\newblock Magnetic monopoles in unified theories,
\newblock Nucl. Phys. B {\bf 79}(CERN-TH-1876), 276--284 (1974).

\bibitem{polyakov1974}
A.~M. Polyakov,
\newblock Spectrum of particles in quantum field theory,
\newblock JETP Lett {\bf 20}, 430--433 (1974).

\bibitem{goddard1978}
P.~Goddard and D.~. Olive,
\newblock Magnetic monopoles in gauge field theories,
\newblock Reports on Progress in Physics {\bf 41}(9), 1357 (1978).

\bibitem{weinberg2012classical}
E.~J. Weinberg,
\newblock {\em Classical solutions in quantum field theory: Solitons and
  Instantons in High Energy Physics},
\newblock Cambridge University Press, 2012.

\bibitem{montonen1977magnetic}
C.~Montonen and D.~I. Olive,
\newblock Magnetic monopoles as gauge particles?,
\newblock Phys. lett. B {\bf 72}(CERN-TH-2391), 117--120 (1977).

\bibitem{fring2019goldstone1}
A.~Fring and T.~Taira,
\newblock Goldstone bosons in different PT-regimes of non-Hermitian scalar
  quantum field theories,
\newblock Nucl. Phys. B {\bf 950}, 114834 (2020).

\bibitem{fring2019goldstone2}
A.~Fring and T.~Taira,
\newblock Pseudo-Hermitian approach to Goldstone’s theorem in non-Abelian
  non-Hermitian quantum field theories,
\newblock Phys. Rev. D {\bf 101}(4), 045014 (2020).

\bibitem{fring2020higgs}
A.~Fring and T.~Taira,
\newblock Massive gauge particles versus Goldstone bosons in non-Hermitian
  non-Abelian gauge theory,
\newblock arXiv preprint arXiv:2004.00723  (2020).

\bibitem{kings_eqm1}
J.~Alexandre, P.~Millington, and D.~Seynaeve,
\newblock Symmetries and conservation laws in non-Hermitian field theories,
\newblock Phys. Rev. D {\bf 96}(6), 065027 (2017).

\bibitem{kings_eqm2}
J.~Alexandre, D.~Seynaeve, and P.~Millington,
\newblock Consistent description of field theories with non-Hermitian mass
  terms,
\newblock in {\em J. Phys. Conf. Ser.}, volume 952, page 012012, 2017.

\bibitem{kings_abelian_higgs}
J.~Alexandre, J.~Ellis, P.~Millington, and D.~Seynaeve,
\newblock Gauge invariance and the Englert-Brout-Higgs mechanism in
  non-Hermitian field theories,
\newblock Phys. Rev. D {\bf 99}(7), 075024 (2019).

\bibitem{kings_goldstone}
J.~Alexandre, J.~Ellis, P.~Millington, and D.~Seynaeve,
\newblock Spontaneous symmetry breaking and the Goldstone theorem in
  non-Hermitian field theories,
\newblock Phys. Rev. D {\bf 98}(4), 045001 (2018).

\bibitem{kings_non-abelian_higgs}
J.~Alexandre, J.~Ellis, P.~Millington, and D.~Seynaeve,
\newblock Spontaneously breaking non-Abelian gauge symmetry in non-Hermitian
  field theories,
\newblock Phys. Rev. D {\bf 101}(3), 035008 (2020).

\bibitem{mannheim2019goldstone}
P.~D. Mannheim,
\newblock Goldstone bosons and the Englert-Brout-Higgs mechanism in
  non-Hermitian theories,
\newblock Phys. Rev. D {\bf 99}(4), 045006 (2019).

\bibitem{bogomol1976stability}
E.~B. Bogomolny,
\newblock The stability of classical solutions,
\newblock Sov. J. Nucl. Phys.(Engl. Transl.);(United States) {\bf 24}(4)
  (1976).

\bibitem{prasad1975exact}
M.~K. Prasad and C.~M. Sommerfield,
\newblock Exact classical solution for the't Hooft monopole and the Julia-Zee
  dyon,
\newblock Phys. Rev. Lett {\bf 35}(12), 760 (1975).

\bibitem{polyakov1996particle}
A.~M. Polyakov,
\newblock Particle spectrum in quantum field theory,
\newblock in {\em 30 Years Of The Landau Institute—Selected Papers}, pages
  540--541, World Scientific, 1996.

\bibitem{ptbook}
C.~M. Bender,
\newblock {\em PT symmetry: In quantum and classical physics},
\newblock World Scientific Publishing, 2018.

\bibitem{mostafazadeh2010pseudo}
A.~Mostafazadeh,
\newblock Pseudo-Hermitian representation of quantum mechanics,
\newblock International Journal of Geometric Methods in Modern Physics {\bf
  7}(07), 1191--1306 (2010).

\bibitem{bender2005dual}
C.~M. Bender, H.~F. Jones, and R.~J. Rivers,
\newblock Dual PT-symmetric quantum field theories,
\newblock Phys. Lett. B {\bf 625}(3-4), 333--340 (2005).

\bibitem{derrick_scaling_argument}
G.~H. Derrick,
\newblock Comments on nonlinear wave equations as models for elementary
  particles,
\newblock J. Math. Phys {\bf 5}(9), 1252--1254 (1964).

\bibitem{general_solution_to_higgs_vacuum}
E.~Corrigan, D.~I. Olive, J.~Nuyts, and D.~B. Fairlie,
\newblock Magnetic monopoles in SU (3) gauge theories,
\newblock Nucl. Phys. {\bf 106}(CERN-TH-2102), 475--492 (1975).

\bibitem{first_mensioned_higgs_vacuum}
P.~Goddard and D.~I. Olive,
\newblock Magnetic monopoles in gauge field theories,
\newblock Rep. Prog. Phys {\bf 41}(9), 1357 (1978).

\bibitem{Reference_for_winding_number}
J.~Arafune, P.~G.~O. Freund, and C.~J. Goebel,
\newblock Topology of Higgs fields,
\newblock J. Math. Phys {\bf 16}(2), 433--437 (1975).

\bibitem{physical_motivation_for_BPS}
T.~W. Kirkman and C.~K. Zachos,
\newblock Asymptotic analysis of the monopole structure,
\newblock Phys. Rev. D {\bf 24}(4), 999 (1981).

\end{thebibliography}

\end{document}